\documentclass[conference]{IEEEtran}
\IEEEoverridecommandlockouts
\usepackage{cite}
\usepackage{amsmath,amssymb,amsfonts}
\usepackage{algorithmic}
\usepackage{graphicx}
\usepackage{siunitx}
\usepackage{textcomp}
\usepackage{xcolor}
\def\BibTeX{{\rm B\kern-.05em{\sc i\kern-.025em b}\kern-.08em
    T\kern-.1667em\lower.7ex\hbox{E}\kern-.125emX}}
\DeclareMathOperator{\diag}{diag}

\begin{document}

\title{nRTIS: Low-Cost Real-Time 3D Sonar Imaging Circular Array Supporting Beamforming for Industrial
Applications}

\author{\IEEEauthorblockN{
Rens Baeyens\IEEEauthorrefmark{1}\IEEEauthorrefmark{2},
Dennis Laurijssen\IEEEauthorrefmark{1}\IEEEauthorrefmark{2},
Jan Steckel\IEEEauthorrefmark{1}\IEEEauthorrefmark{2} and
Walter Daems\IEEEauthorrefmark{1}\IEEEauthorrefmark{2}}
\IEEEauthorblockA{\IEEEauthorrefmark{1}FTI Cosys-Lab,
University of Antwerp,
Antwerp, Belgium, Email: walter.daems@uantwerpen.be}
\IEEEauthorblockA{\IEEEauthorrefmark{2}Flanders Make Strategic Research Centre
Lommel, Belgium}}

\maketitle

\begin{abstract}
Conventional ultrasonic inspection systems rely on phased arrays and high-performance computing hardware, making them costly, bulky, and unsuitable for portable or embedded use. In this work, we present nRTIS (nano Real-Time 3D Imaging Sonar), a compact ultrasonic sensing platform built around a circular array of MEMS microphones and a central ultrasonic transducer. The device achieves real-time acquisition through an RP2350 microcontroller and high-speed USB transfer. We validate the system using both simulations and controlled experiments: point spread function (PSF) simulations demonstrate beamforming resolution and sidelobe suppression, while reflector measurements confirm robust data acquisition. These results highlight the potential of nRTIS for scalable industrial applications such as weld inspection, pipe mapping, and robotic navigation.
\end{abstract}


\section{Introduction}
Non-destructive evaluation (NDE) techniques play a critical role in industrial settings by enabling the inspection of materials and structures without compromising their integrity~\cite{Abbasi2022, Malyy2022, Athi2009, Gomes2024, Yildiz2019, Li2024, Abaravicius2025, Schenck2025,Kerstens2019,Izquierdo2024, Ziaja-Sujdak2025,Qi2024}. Ultrasonic imaging is particularly valued for its ability to detect subsurface features with high spatial resolution. However, conventional ultrasonic systems rely on expensive phased arrays and high-throughput computing hardware, making them large, costly, and power-hungry~\cite{Haugwitz2024,Allevato2022, Allevato2021,Laurijssen2024}. These traits severely limit their deployment in mobile platforms or confined environments.

Advances in MEMS (Microelectromechanical Systems) microphones and embedded microcontrollers have enabled a new class of miniaturized acoustic sensing systems~\cite{Kerstens2023,Verellen2020,Haugwitz2022}. These offer the potential for scalable, low-cost sensing, in portable and embedded formats. Yet, delivering real-time, high-resolution 3D imaging at such small scales remains challenging, especially when robust beamforming and noise suppression are required.

This work introduces nRTIS (nano-scale Real-Time Imaging Sonar), a compact 3D ultrasonic sensing system with a total aperture diameter of just \qty{30}{\milli\meter}. The sensor integrates a circular array of 16 MEMS microphones surrounding a central ultrasonic transducer, controlled by an RP2350 dual-core microcontroller. While the embedded hardware handles real-time ultrasonic data acquisition and streaming, advanced spatial filtering is performed externally on a host system.
By offloading beamforming computations, nRTIS achieves the benefits of high-resolution 3D acoustic imaging while maintaining an extremely small physical and computational footprint. The system’s \qty{30}{\milli\meter} form factor enables integration into handheld tools, mobile robots, and inspection probes: applications where traditional ultrasonic equipment is too large or inflexible. This paper describes the system architecture, acquisition pipeline, and early performance evaluations, and discusses its potential for applications such as weld inspection, pipe mapping, and autonomous navigation in constrained industrial environments. The system is validated via Point Spread Function simulations and a controlled reflector experiment, demonstrating the feasibility of low-cost, compact ultrasonic imaging.

\section{System Architecture}
\label{system_arch}

\subsection{Embedded hardware}
\label{hardware}
\begin{figure}[ht]
\centering
    \includegraphics[width=\linewidth]{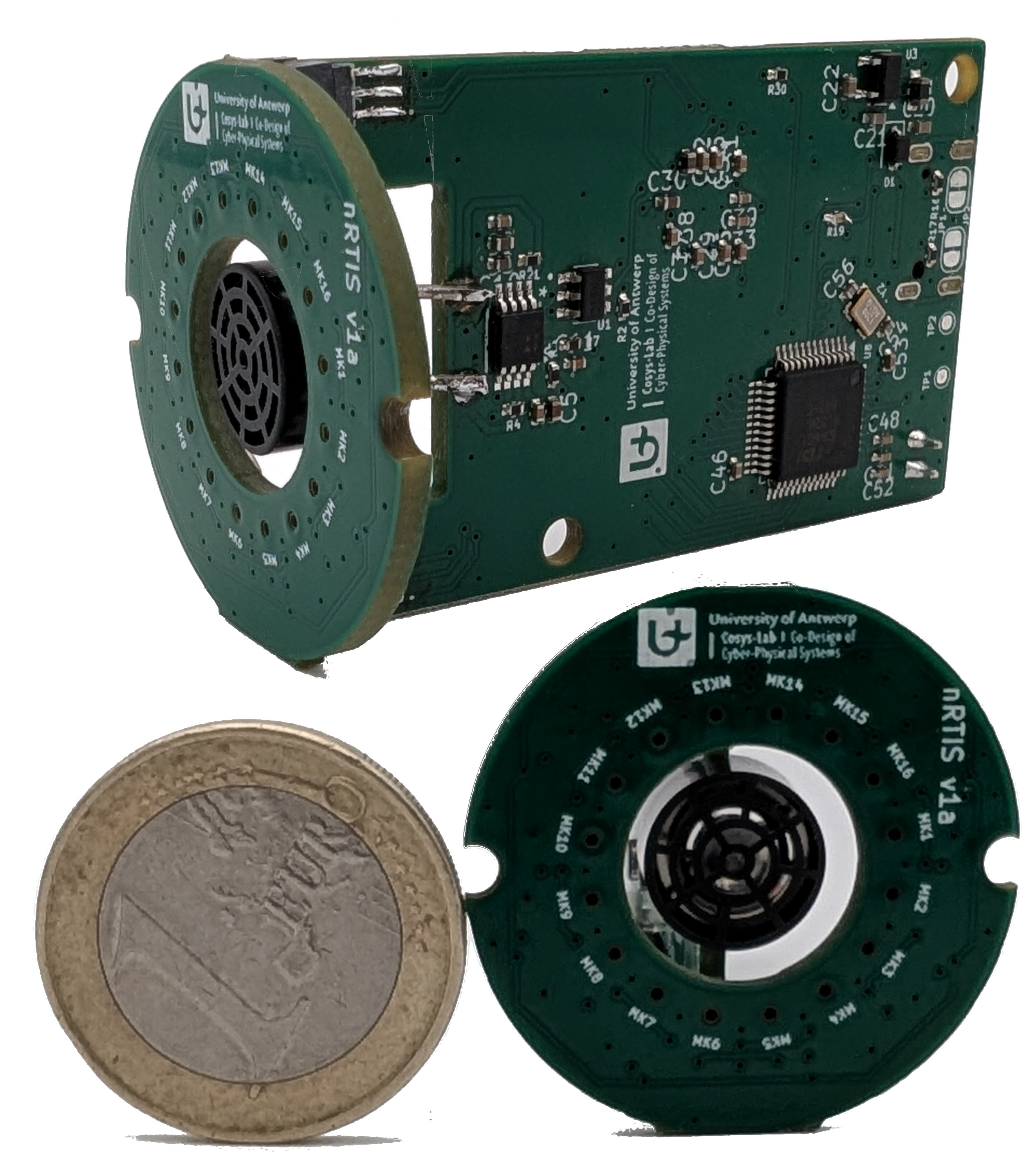}
    \includegraphics[width=\linewidth]{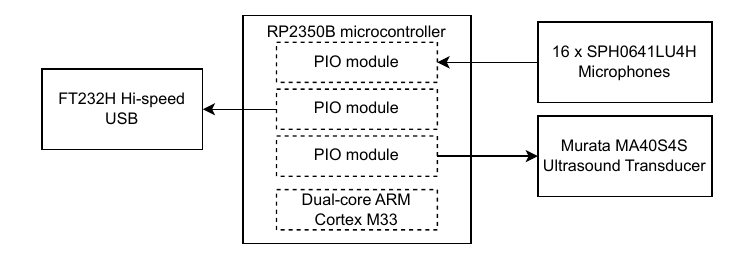}
    \caption{The developed nRTIS hardware platform with the 30mm diameter aperture front-end containing 16 Knowles SPH0641LU4H-1 MEMS microphones and a single Murata MA40S4S ultrasound transducer. The back-end contains the transducer driving circuitry, the RP2350B microcontroller, and an FT232H usb bridge. The aperture has a diameter of $30\unit{\milli\meter}$, matching the form factor of typical one dimensional ultrasound sensors.}
    \label{fig:syshardware}
\end{figure}

The nRTIS front-end features a circular array of 16 Knowles SPH0641LU4H-1 MEMS microphones arranged to provide near-uniform angular coverage. A central Murata MA40S4S ultrasonic transducer with a center frequency of $\qty{40.0}{\kilo\hertz}$ emits short pulses for active probing. The array's compact geometry of $\qty{30}{\milli\meter}$ diameter enables integration into robotic arms, handheld probes, or confined inspection systems. The back-end includes the RP2350 microcontroller and an FT232H High-speed USB bridge to enable data transfer up to $\qty{480}{Mb\per\second}$. Figure~\ref{fig:syshardware} illustrates the complete hardware platform.

\subsection{Firmware and Data Acquisition}
\label{firmware}
The RP2350 microcontroller is used to control both acquisition and transmission. Transmission is handled by a single PIO module that communicates a predefined audio sequence to a Digital Analog Converter (DAC) at fixed intervals. Real-time acquisition is realised using two PIO (Programmable I/O) modules. The microphones' pulse density modulated (PDM) signals are sampled synchronously at $\qty{4.45}{\mega\hertz}$ on one PIO, where acoustic frames are buffered and transferred to a second PIO using DMA. The second PIO controls an FT232H USB driver to transfer the acquired data in real-time to a remote host. This pipeline enables real-time streaming. The main controller only handles the timed triggers of the transmission, and the trigger of the DMA transfer of buffered acquisition frames between the two PIO modules. The main cores remain available for future on-device pre-processing.

\section{Signal Processing}
\subsection{Signal Model}
Considering an $L$ elements microphone array and denoting the complex envelope of the received signals as $y[t] \in \mathbb{C}^{L\times1}$. The received signals are related to the transmitted signal by:
\begin{equation}
    y(t) = d(t)s(t) + G(t)s_I(t) + v(t)
\end{equation}
where the desired signal is denoted $s$, its power is $\sigma_{d}^{2}$, and d is the corresponding steering vector. $s_{I} =[ s_{I,1},\dots,s_{I,K}]^T$ is a vector of K uncorrelated interference signals, $\Lambda = \diag([\sigma^{2}_{I,1},\dots,\sigma^{2}_{I,K}]$ is its corresponding correlation matrix, and $G = [g_{1}, \dots, g_{K}]$ is a matrix consisting of the corresponding steering vectors. $v$ are spatially-white sensor noise signals with correlation matrix $\sigma_{v}^{2}I$. 
The correlation matrix of the received signal $y$ can be given by 
\begin{equation}
    R_y = R_s + R_n
\end{equation}
where $R_s = \sigma_{d}^{2}dd^{\mathrm{H}}$ is the desired signal component, and $R_n = G\Lambda G^{\mathrm{H}} + \sigma_v^2 I$ is the interference signals component. Defining $B=G\Lambda^{\frac{1}{2}}$ the interference signals correlation matrix can be restated as $R_n = BB^{\mathrm{H}} + \sigma^2_v I$.

The received signal is processed by a beamformer, in order to enhance the desired signal while mitigating interference signals: 
\begin{equation}
    x = w^{\mathrm{H}}y.
\end{equation}

Then one can estimate the direction of arrival of the target signal by scanning the power spectrum in a grid using the steering vector $d$ and search for beam peaks. 

\section{Preliminary Results}
\subsection{Simulation of Point Spread Functions}
\begin{figure}[hb]
\centering
    \includegraphics[width=\linewidth]{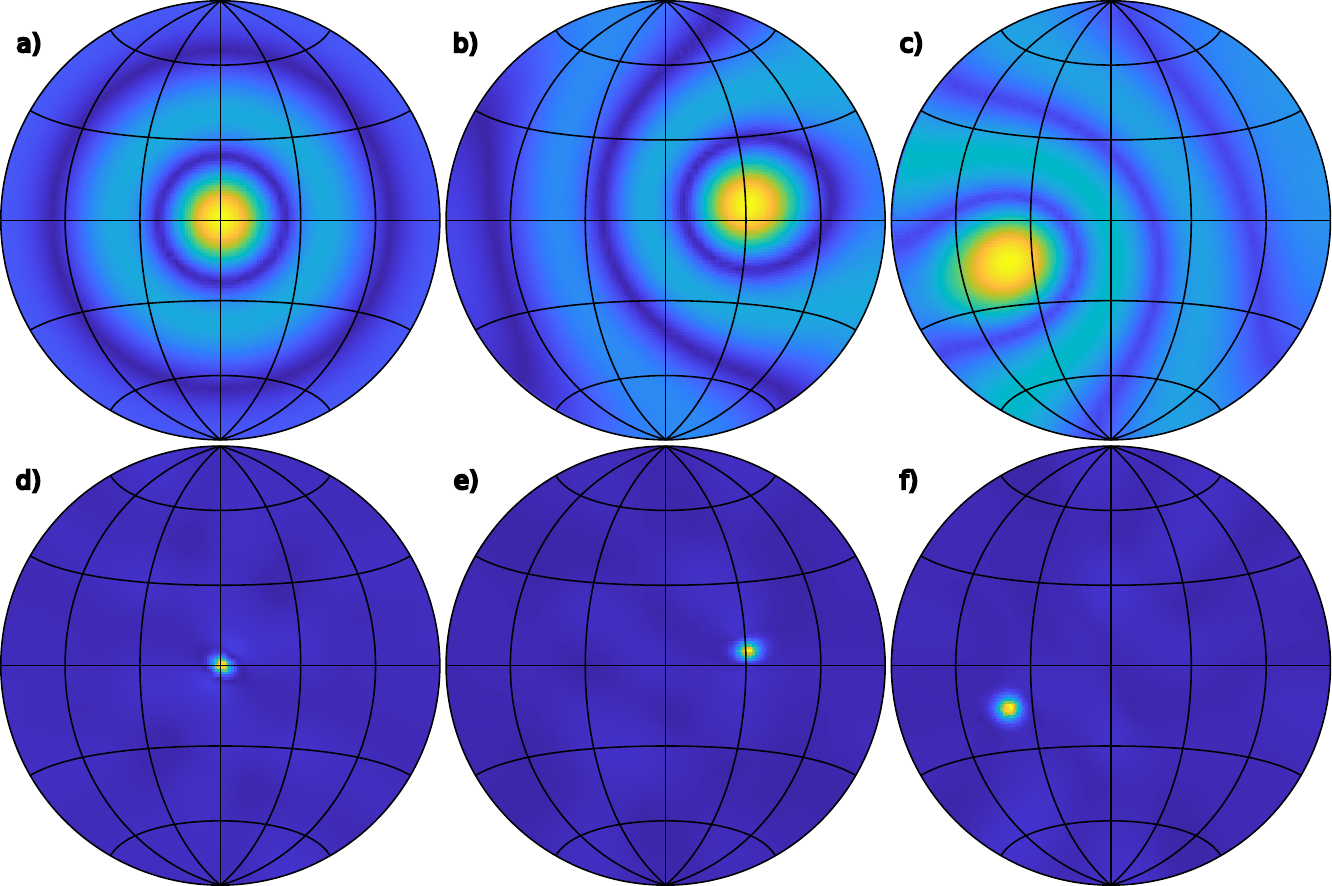}
    \caption{Point Spread Functions of point sources placed at different spatial locations: panel a \& d) $(\theta,\phi) = (\qty{0}{\degree},\qty{0}{\degree})$, panel b \& e) $(\theta,\phi) = (\qty{30}{\degree},\qty{0}{\degree})$, and panel c \& f) $(\theta,\phi) = (\qty{-45}{\degree},\qty{-15}{\degree})$
Panels a-c) show the response of the system using Bartlett beamforming, and panels d-f) show the response when using the MVDR beamformer. The PSFs are shown on a logarithmic scale}
    \label{fig:PSF}
\end{figure}

In order to verify the functionality of the nRTIS sensor, a simulation model of the sensor was developed, following the equations derived in \cite{VanTrees2002}. We calculate a so-called Point Spread Function (PSF) \cite{Steckel2015,Kerstens2019} of the sensor system, which describes the image obtained by the sensor in response to a Dirac-like point source in space. We placed the point source in three spatial locations, defined by their azimuth angle ($\theta$) and elevation angle ($\phi$): $(\theta,\phi) = (\qty{0}{\degree},\qty{0}{\degree}), (\qty{30}{\degree},\qty{0}{\degree})$ and $(\theta,\phi) = (\qty{-45}{\degree},\qty{-15}{\degree})$. The resulting PSFs can be found in Figure~\ref{fig:PSF}, with panels a–c presenting the PSF calculated using conventional Bartlett beamforming, and panels d–f presenting the PSF using MVDR beamforming.

Compared to Bartlett, the MVDR beamformer produces a PSF with a significantly narrower main lobe, which improves angular resolution and the ability to distinguish between closely spaced sources. At the same time, the sidelobe levels are strongly suppressed, reducing the level of spurious responses and thus mitigating the risk of false detections.

\subsection{Real-world validation: active measurement}
\begin{figure}[t]
\centering
    \includegraphics[width=\linewidth]{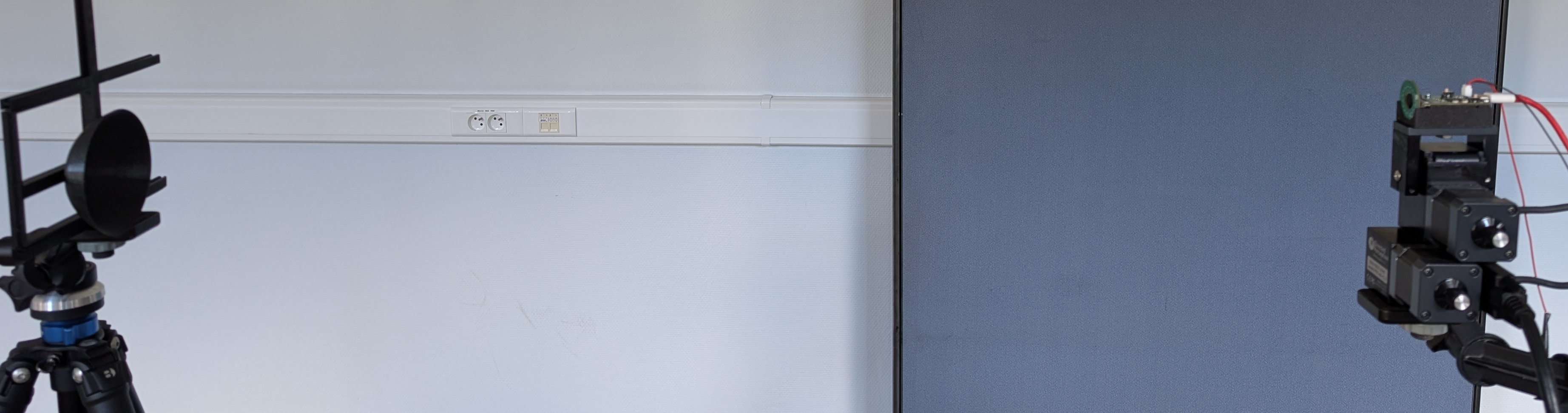}
    \caption{Experimental setup: the 3D-printed spherical reflector (radius $\qty{55}{\milli\meter}$) positioned $\qty{1}{\meter}$ from the nRTIS, which is mounted on a pan–tilt system for azimuthal scanning.}
    \label{fig:pantilt}
\end{figure}

\begin{figure}[h]
\centering
    \includegraphics[width=\linewidth]{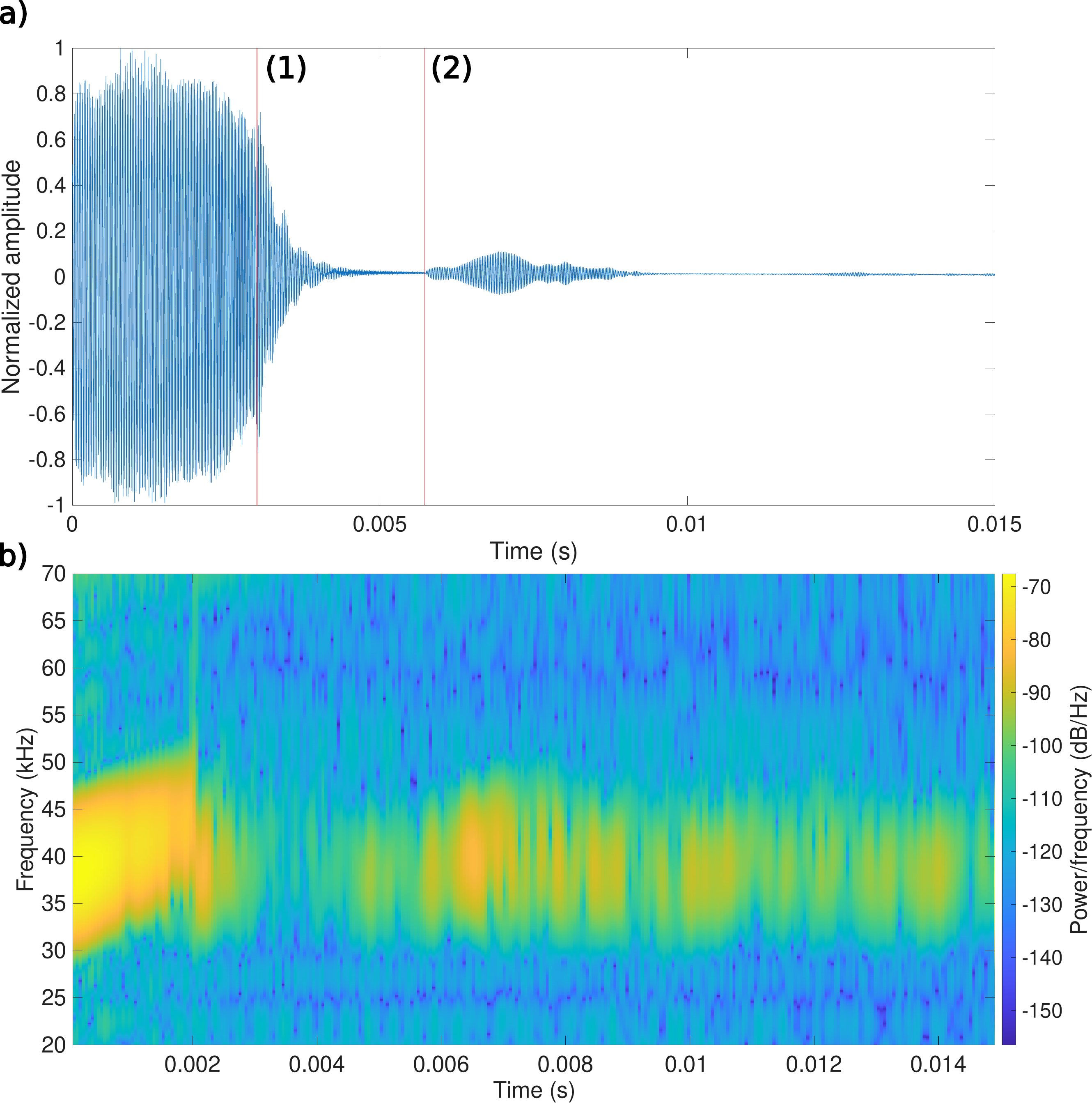}
    \caption{Experimental results: the transmitted linear sweep and the resulting echo from the reflector placed at one meter. (1) indicates the end of linear sweep transmission, whereas (2) indicates the expected start of the reflector echo.}
    \label{fig:transmit_result}
\end{figure}

To evaluate the range functionality and acquisition accuracy of the nRTIS system, a controlled measurement setup was constructed as presented in Figure~\ref{fig:pantilt}. As a reference target, a 3D-printed spherical reflector with a radius of $\qty{55}{\milli\meter}$, following the design in \cite{Jansen2024}, was employed. The center of the reflector was positioned at a distance of $\qty{1}{\meter}$ from the nRTIS aperture when aligned with $\qty{0}{\degree}$ incidence. The nRTIS system itself was mounted on a motorized pan–tilt stage, enabling systematic variation of the relative orientation between the sensor and the reflector. A continuous measurement of $\qty{3}{\second}$ was recorded while actively probing at $\qty{10}{\hertz}$ with a $\qty{3}{\milli\second}$ linear frequency sweep ranging from $\qty{36}{\kilo\hertz}$ to $\qty{44}{\kilo\hertz}$. The resulting emitted signal and received echo are presented in Figure~\ref{fig:transmit_result}.

\section{Discussion \& Future Work}
The point spread function simulations and reflector experiments together demonstrate the feasibility of compact MEMS-based ultrasonic imaging. MVDR beamforming improves angular resolution and reduces sidelobes, while the reflector experiments validate the practical acquisition pipeline and confirm that echoes can be reliably detected in a controlled setting.

Future work will focus on extending the system’s capabilities by implementing adaptive beamforming techniques with spatial smoothing to improve robustness against multipath reflections and correlated interference. Additionally, spatial localization algorithms will be developed to enable 3D positioning of reflectors, supporting applications such as weld defect mapping, pipe reconstruction, and robotic navigation. Onboard pre-processing and partial beamforming on the RP2350 will be explored to reduce the dependence on host-side computation and enable more mobile operation. Finally, the system will be tested in realistic industrial environments, including metallic pipes, weld seams, and cluttered factory settings, to validate its practical performance.

\section{Conclusion}
This work presented nRTIS, a compact real-time 3D sonar system based on MEMS microphones and an RP2350 microcontroller. Simulations of point spread functions showed that adaptive beamforming can significantly enhance angular resolution and suppress sidelobes, while controlled experiments with a spherical reflector demonstrated reliable echo detection at one meter. The combination of simulation and experimental validation highlights both the potential and current limitations of the system. By addressing robustness, on-device processing, and operational range, nRTIS has the potential to become a versatile platform for industrial inspection, robotic navigation, and ultrasonic imaging in scenarios where conventional systems are impractical.

\bibliographystyle{IEEEtran}
\bibliography{fulllib}

\end{document}